\documentclass[pre,twocolumn,showpacs]{revtex4}
\usepackage{graphicx}

\def\be{\begin{equation}}
\def\ee{\end{equation}}

\begin{document}
\title{The PDF of fluid particle acceleration in turbulent flow
with underlying normal distribution of velocity fluctuations}
 \author{A.K. Aringazin}
 \email{aringazin@mail.kz}
  \altaffiliation[Also at ]{Kazakhstan Division, Moscow State
University, Moscow 119899, Russia.}
 \affiliation{Department of Theoretical Physics, Institute for
Basic Research, Eurasian National University, Astana 473021
Kazakhstan}

 \author{M.I. Mazhitov}
 \email{mmi@emu.kz}
 \affiliation{Department of Theoretical Physics, Institute for
Basic Research, Eurasian National University, Astana 473021
Kazakhstan}

\date{15 May 2003}

\begin{abstract}
We describe a formal procedure to obtain and specify the general
form of a marginal distribution for the Lagrangian acceleration of
fluid particle in developed turbulent flow using Langevin type
equation and the assumption that velocity fluctuation $u$ follows
a normal distribution with zero mean, in accord to the
Heisenberg-Yaglom picture. For a particular representation,
$\beta=\exp[u]$, of the fluctuating parameter $\beta$, we
reproduce the underlying log-normal distribution and the
associated marginal distribution, which was found to be in a very
good agreement with the new experimental data by Crawford,
Mordant, and Bodenschatz on the acceleration statistics. We
discuss on arising possibilities to make refinements of the
log-normal model.
\end{abstract}

\pacs{05.20.Jj, 47.27.Jv \hfill [To appear in Phys. Lett. A
(2003)]}

\maketitle

\section{Introduction}

In the context of Tsallis formalism \cite{Tsallis} and generalized
statistics approach \cite{Johal,Aringazin,Beck3} to developed
turbulence, C.~Beck \cite{Beck4} used the underlying log-normal
distribution, $f(\beta)$, to describe fluctuation of the parameter
$\beta$ entering some Langevin equation for the acceleration $a$
of fluid particle in the Lagrangian frame; see
also~\cite{Beck2,Wilk}. For the case of a linear drift force, the
resulting marginal probability density function of the
acceleration is \cite{Beck4}
\be\label{1}
P(a) = \frac{1}{2\pi s}\int_{0}^{\infty}\! d\beta\
\beta^{-1/2}
\exp\left[-\frac{(\mathrm{ln}\frac{\beta}{m})^2}{2s^2}\right]
e^{-\frac{1}{2}\beta a^2},
\ee
where $m=\exp[s^2]$ provides a unit variance and $s$ is a fitting
parameter, $s^2=3.0$. This distribution was found to be in a very
good agreement with the recent Lagrangian experimental data by
Porta, Voth, Crawford, Alexander, and Bodenschatz
\cite{Bodenschatz}, the new data by Crawford, Mordant, and
Bodenschatz (the Taylor microscale Reynolds number is
$R_\lambda=690$, the normalized acceleration range is $[-60,60]\ni
a$, the Kolmogorov timescale $\tau_\eta$ is
resolved)~\cite{Bodenschatz2}, Mordant, Delour, Leveque, Arneodo,
and Pinton ($R_\lambda=740$, $a\in [-20,20]$, $\tau_\eta$ is not
resolved) \cite{Mordant}, and direct numerical simulations of the
Navier-Stokes equations by Kraichnan and Gotoh ($R_\lambda=380$,
$a\in [-150,150]$)~\cite{Gotoh}.

The same approach with the underlying $\chi^2$ distribution,
$f(\beta)$, of fluctuations resulted in an analytically explicit
distribution \cite{Aringazin3},
\be\label{2}
P(a) =
\frac{C\exp[-a^2/a_c^2]}{(1+\frac{1}{2}V_0(q-1)a^2)^{1/(q-1)}},
\ee
where $C$ is a normalization constant, $V_0=4$, and $q=3/2$
(Tsallis entropic index) are due to the theory, and $a_c$ is a
free parameter used for a fitting, $a_c=36.0$. This distribution
was found to be in a good agreement with the experiments
\cite{Bodenschatz2}. The result (\ref{2}) is based on the
consideration of Ref.~\cite{Beck}.

In general, the marginal distribution is defined as
\be\label{3}
P(a) = \int_{0}^{\infty}\! d\beta\ P(a|\beta)f(\beta),
\ee
where $P(a|\beta)$ is a probability density function associated to
a surrogate dynamical equation, the Langevin equation for the
acceleration $a$~\cite{Beck},
\be\label{4}
\partial_t a = \gamma F(a) + \sigma L(t).
\ee
Here, $F(a)$ is a "drift force", $L(t)$ is delta-correlated
Langevin source (Gaussian white noise, $\langle L(t)\rangle=0$,
$\langle L(t)L(t')\rangle =2\delta(t-t')$),
$\beta=\gamma/\sigma^2$ is assumed to be a fluctuating real
positive parameter. For constant parameters $\gamma$ and $\sigma$,
this model assumes that the stochastic process (\ref{4}) is
Markovian, and $P(a|\beta)$ is found as a stationary solution of
the associated Fokker-Planck equation. For a linear drift force,
$F(a)=-a$, the stationary conditional distribution, $P(a|\beta)$,
is found to be
\be\label{5}
P(a|\beta)= C(\beta)\exp[-\beta a^2/2],
\ee
where $C(\beta)$ is a normalization constant, $a\in
[-\infty,\infty]$.

The function $f(\beta)$ entering Eq.~({3}) is probability density
function arising from the assumption that the parameter $\beta$ is
a stochastic variable. For a constant $\beta$, the Gaussian
solution (\ref{5}) meets the large scale (large time increment)
statistics but it fails to describe observed Reynolds number
dependent stretched exponential tails of the experimental small
scale $P(a)$ \cite{Bodenschatz} that correspond to high
probability to find extremely big values of the fluid particle
acceleration in the developed turbulent flow.

The interest in studying Langevin type equations to describe
developed turbulence is motivated by the recent high precision
Lagrangian experiments \cite{Bodenschatz,Bodenschatz2,Mordant},
which give an important dynamical information and new look to the
intermittency in fluid turbulence. Time response characteristics
of the tracer polystyrene particle and the precision in the
experiments \cite{Bodenschatz,Bodenschatz2} allow to resolve about
1/20 of the Kolmogorov time and 1/40 of the Kolmogorov length
($R_\lambda=970$) so that the acceleration is really resolved and
particle follows rare violent events (within 7\% of the ideal
value of acceleration even at the high Reynolds numbers studied
there), with the collected statistical data sufficient to
establish convergence of the fourth moment of acceleration.

In Sec.~2, we construct a simple model that allows one to derive
the marginal probability density function, $P(a)$, of the
acceleration of fluid particle in turbulent flow, with underlying
normally distributed velocity fluctuations. In Sec.~3, we
reproduce the result of Ref.~\cite{Beck4} as a particular case,
and propose a generalization of the log-normal model. In Sec.~4,
we briefly summarize the results and make a few remarks.

\section{The marginal distribution with underlying normally
distributed velocity fluctuations}

While the conditional distribution $P(a|\beta)$ can be found
starting from the Langevin equation (\ref{4}), we do not have
sufficiently strong theoretical requirements to determine a unique
form of the distribution $f(\beta)$ except for that it should obey
general conditions, such as that the integral in Eq.~(\ref{3})
should converge and the resulting marginal distribution $P(a)$
should be normalizable. To select a particular form of the
distribution $f(\beta)$, one can use {\it ad~hoc} statistical
distributions or make assumptions stemming from considerations of
the developed turbulence.

As a first step, one can associate the parameter $\beta$ with the
mean energy dissipation rate $\epsilon$, and use the relationship
of $\epsilon$ with the fluid velocity $v_i$, within the framework
of Kolmogorov scaling theory of developed turbulence.

Second, as shown by Renner, Peinke, and
Fried\-rich~\cite{Friedrich}, the small scale intermittency can be
traced back to a stochastic nature of the averaged energy
dissipation rate \cite{Landau}, and the Markovian condition for
the stochastic process is fulfilled at scales larger than the
Taylor microscale $\lambda$. Thus, a stochastic nature of $\beta$
can be related to the stochastic energy dissipation rate (measured
in Eulerian experiments), which we denote by $\tilde\epsilon$.

For example, the assumptions $\beta \sim \tilde\epsilon$ and
$\tilde\epsilon \sim \sum u_i^2$, discussed at length by Beck
\cite{Beck} (we do not repeat it here for brevity), for the
fluctuating, averaged over a ball of radius $r$, energy
dissipation $\tilde\epsilon$ and velocity fluctuations $u_i$ at
the Kolmogorov scale were used to propose $\beta \sim \sum u_i^2$
and to select the $\chi^2$ distribution of $\beta$.
Three-dimensionality of the space was used explicitly to determine
the value of the free parameter (Tsallis entropic index) in this
model. A comparison of the formalism based on Eq.~(\ref{4}) with
the Sawford model \cite{Sawford} implies $\beta =
2C_0^{-2}a_0\nu^{1/2}\epsilon^{-3/2}$ \cite{Beck4} so that with
the replacement of $\epsilon$ by the fluctuating $\tilde\epsilon$
and using again $\tilde\epsilon \sim \sum u_i^2$, one can obtain a
different relation, $\beta \sim \sum u_i^{-3}$. The Kolmogorov's
1941 approach implies $\epsilon \sim {\bar u}^3$, where $\bar u$
stands for the {\em rms} velocity, which can also be used to try
to relate $\tilde\epsilon$ to the velocity fluctuations. Although
successful in capturing main features of the experimental data,
different models use different powers of $\epsilon$ to represent
$\beta$ that makes a theoretical problem when selecting an
appropriate distribution of $\beta$.

It is however common to the above representations that velocity
fluctuations, $u_i$, are assumed to be normally distributed with
zero mean, and that statistics of $\beta$, $\tilde\epsilon$, and
$u_i$ are related to each other due to some functional
dependencies between their characteristics holding in the inertial
range. A direct dependence of {\em statistical} properties of the
acceleration $a$ on velocity fluctuations was established in the
wellknown Heisenberg-Yaglom theory. This gives a different look to
the problem since the Lagrangian description refers to individual
trajectories of fluid particles as compared with the well
established Eulerian framework, in which the intermittency is
understood in terms of anomalous scaling of the moments of
velocity increments in space related to the stochastic nature of
the energy dissipation rate (non-dynamical description). From a
simplified dynamical point of view, one can think of that the
tracer particle trajectory extends to a large region (few integral
length scales, in the experiments) crossing during the course
subregions characterized by different {\em local} amplitudes of
velocity fluctuations which are randomly distributed in space and
varies with time (Brownian like motion of the particle driven by
the stochastic delta-correlated force with stochastically slow
varying intensity). The Lagrangian velocity autocorrelation
function is known to decay very slowly, to vanish at times bigger
than the integral time scale. This view is valid for sufficiently
{\em high} Reynolds numbers and meets that provided by the
stochastic energy dissipation rate argued earlier~\cite{Beck}.
This is also in an agreement with the remark made in
Ref.~\cite{Bodenschatz} that extremely high accelerations seem to
be associated with coherent structures (geometrically these may be
thought of as very thin tornadoes) which persist for many
Kolmogorov times, substantially longer than the correlation time
of the acceleration components. At lower Reynolds numbers,
$R_\lambda<500$, the acceleration is increasingly coupled to large
scales of the flow in a dynamical way so that a different approach
may be required to describe such a situation to a good accuracy.

In the Lagrangian frame, it is then natural to relate {parameters}
of the stochastic dynamical equation (\ref{4}), which describes
the acceleration in the context of a generalized Brownian motion,
to velocity fluctuations due to the Heisenberg-Yaglom picture of
developed turbulence. Ultimately of course we deal with a {\em
statistical} description of the acceleration and the velocity
fluctuations within the framework of Fokker-Planck equation
associated to Eq.~(\ref{4}). Guided by this observation, one can
develop a general consideration along this line of reasoning.

In the present paper, we formulate some general requirements for
$f(\beta)$ using the assumption that the velocity fluctuations
$u_i$ are Gaussian distributed random variables with zero means,
$\langle u_i\rangle =0$, and variances $s_i$,
\be\label{6}
g(u_i) = C(s_i) \exp\left[-{u_i^2}/{(2s_i^2)}\right],
\ee
where $C(s_i)$ is normalization constant; $u_i \in
[-\infty,\infty]$, $i=1,2,3$. This assumption meets the Lagrangian
experimental data~\cite{Bodenschatz}, for each component of the
velocity fluctuations.

It should be stressed that our treatment is made regardless to a
particular functional dependence of $\beta$ on $\tilde\epsilon$,
or a particular form of $f(\beta)$, and provides a simple way to
formalize the general model in accord to the Heisenberg-Yaglom
picture, which relates some statistical properties of the
acceleration to that of the velocity fluctuations.

We assume that $\beta$ depends on the velocity $v_i$ the dynamics
of which is decoupled from that of the acceleration so that the
well known fluctuating character of $v_i$ for the turbulent flow
determines fluctuations of $\beta$. Below we drop the index $i$ to
simplify notation. Also, we treat the developed turbulent flow to
be statistically isotropic and translation invariant thus
discarding skewness effects; the statistical anisotropy is known
to become very small with the increase of the Reynolds number,
e.g., at $R_\lambda > 900$.

In the mean field approximation, we can represent $v =
\langle v \rangle + u$, where the fluctuation $u$ is
characterized by zero mean, $\langle u \rangle =0$. Hence, we can
write
\be\label{Eq:beta}
\beta = \beta_c(\langle v \rangle)+B(\langle v \rangle,u),
\ee
where we formally separated out $\beta_c$ which depends only on
$\langle v \rangle$.

We suppose that the function $B(u)\equiv B(\langle v \rangle,u)$
satisfies the following natural requirements: (i) It is a
sufficiently smooth invertible function with respect to $u$ so
that the inverse function $u(\beta-\beta_c)$ can be found; (ii) it
maps the interval $[-\infty,\infty]$ to $[-\beta_c,\infty]$, to
provide positiveness of $\beta$; also one can restrict
consideration by the requirement that (iii) there is one-to-one
correspondence between $\beta$ and $u$.

In general, for $\beta$ to be a stochastic variable the function
$B(u)$ should be Borel function of the stochastic variable $u$. We
assume that $u$ has an absolutely continuous distribution function
and the associated probability density function, in particular,
that given by Eq.~(\ref{6}). Not any Borel function $B(u)$
provides existence of a probability density function for $\beta$.
However, a probability density function for $\beta$ {\it a priori}
exists if $B(u)$ is a monotonic function.

Using the general relation
\be\label{Eq:gdu}
g(u)du = g(u(\beta-\beta_c))\left|\frac{du}{d\beta}\right|d\beta
\ee
and Eq.~(\ref{6}), we can make the identification
$f(\beta)=g(u(\beta-\beta_c))|du/d\beta|$, {\it i.e.},
\be\label{7}
f(\beta) = C(s)
\exp\left[{-\frac{[u(\beta-\beta_c)]^2}{2s^2}}\right]
\left|\frac{du}{d\beta}\right|,
\ee
where $C(s)$ is a normalization constant. The inverse function
$u(\beta-\beta_c)$ should provide positive definiteness and
normalizability of the above function $f(\beta)$. From
Eq.~(\ref{3}) we then finally have
\be\label{8}
P(a) \!=\! C(s)\!\!\int_{0}^{\infty}\!\!\!\!\!\! d\beta\,
P(a|\beta)
\exp\!\left[{-\frac{[u(\beta-\beta_c)]^2}{2s^2}}\right]\!
\left|\frac{du}{d\beta}\right|.
\ee
This equation allows one to calculate the distribution $P(a)$ for
given stationary solution $P(a|\beta)$ associated to the Langevin
equation (\ref{4}) and the invertible (monotonic) function $B(u)$.
By construction, the structure of this model is such that the
velocity fluctuation $u$, underlying the dynamics, is normally
distributed with zero mean and variance $s$. Note that only the
{\it amplitude} of $u$ contributes $P(a)$.  Eq.~(\ref{8}) can be
viewed as a specific case of the general equation (\ref{3}) which
captures one of the well known features of the turbulent flow, and
thus enables to rule out some {\it ad hoc} types of distribution
of $\beta$.

Note that in Eq.~(\ref{8}) we did not change the variable over
which the integral is performed as compared to Eq.~(\ref{3}). The
main idea was to introduce the variable $u$ which follows Gaussian
distribution and {\em encode} various underlying distributions in
the functional dependence $\beta = B(u)$. This significantly
reduces the class of admissible probability density functions
$f(\beta)$ that we consider as one of the main results of our
approach.

In the next Section, we study a specific choice of $B(u)$ which is
relevant both from the mathematical and physical points of view.

\section{The log-normal model and beyond}

As mentioned by Beck~\cite{Beck4}, selection of the log-normal
distribution $f(\beta)$ provides that for any power law, $\beta
\sim \tilde\epsilon^\kappa$, the function $\ln\beta =
\kappa\ln\tilde\epsilon$ guarantees that the functional form of
the log-normal distribution does not change, which is viewed as a
hint for a physical relevance of a log-normally distributed
$\beta$. Also, it is known that the probability density function
of the averaged energy dissipation rate is log-normal in agreement
with Kolmogorov's assumption in K62 theory.

The exponential dependence,
\be\label{9}
B(u) = e^{u/u_0},
\ee
where $u_0$ is constant, and $\beta_c=0$, provides a relevant
example for which we get
\be\label{Eq:u} u = u_0\ln\beta, \quad
\left|\frac{du}{d\beta}\right|=\frac{|u_0|}{\beta}.
\ee
The distribution (\ref{7}) then becomes the log-normal
distribution, and Eq.~(\ref{8}) for $P(a|\beta)$ given by
Eq.~(\ref{5}) reproduces the marginal distribution (\ref{1}), with
$u_0=1$.

Also, supposing $\beta \sim \tilde\epsilon^\kappa$, we obtain the
relation
\be\label{Eq:kappa}
\tilde\epsilon \sim e^{u/\kappa},
\ee
where we put $u_0=1$ for simplicity. Again, one can see that the
only measurable effect of the parameter $\kappa$ is that it scales
the variance, $s \to s/\kappa$, in Eq.~(\ref{8}).

One can make the following refinement of the dynamical equation
(\ref{4}) using the relation (\ref{9}), which corresponds to the
log-normal model (\ref{1}). The defining feature of the considered
model is the assumption of slow fluctuating character of the
composite parameter $\beta=\gamma/\sigma^2$. It can be easily
shown that a noisy character of the drift coefficient $\gamma$,
with $F(a)=-a$ and delta-correlated Gaussian white noise $\gamma$
with non-zero mean, can be used to describe intermittency effects
(power law tails) because it is related to wellknown random
multiplicative processes extensively studied in the literature.
Below, we focus on a stochastic nature of the additive noise
amplitude $\sigma$ and study its contribution to the intermittency
separately. A joint effect of both the noisy $\gamma$ and random
intensity of the additive noise $\sigma$ is out of scope of the
present paper and can be studied elsewhere.

From Eq.~(\ref{9}) it follows $\sigma^2=e^{-u/u_0}$ so that the
Langevin equation (\ref{4}) becomes
\be\label{10}
\partial_t a = \gamma F(a) + \exp\left[{-{u}/({2u_0})}\right]L(t),
\ee
where the velocity fluctuation $u$ is treated statistically and
follows Gaussian distribution with zero mean, $F(a)=-a$, and
$\gamma$ is taken to be constant to simplify consideration of the
contribution of the additive noise intensity. The Langevin model
(\ref{10}) meets the Lagrangian experiments to a very good
accuracy as it implies the probability density function of the
form (\ref{1}).

Equation~(\ref{10}) assumes that the dynamics of Lagrangian
acceleration $a$ (the acceleration in the comoving frame, $a_i
\equiv\partial_t v_i + v_k\partial_k v_i$), which comes mainly
from {\em small} scales, and that of the fluctuating energy
dissipation rate $\tilde\epsilon$, or the related {\em amplitude}
of velocity fluctuations, $u_i=v_i-\langle v_i\rangle$, which is
associated mainly to {\em large} scales, are weakly related to
each other at high Reynolds numbers, $R_\lambda>500$, so that in
the lowest approximation it is natural to take them independent.
At small time scale, the velocity fluctuation is characterized by
approximately constant amplitude while its directional part
changes wildly (cf.~\cite{Bodenschatz}). However, in the
statistical context, which reflects Lagrangian dynamics at all the
time scales, the acceleration is related to the amplitude of
velocity fluctuation as it follows, for example, from the
Heisenberg-Yaglom scaling, $\langle a^2 \rangle
\simeq a_0{\bar u}^{9/2}\nu^{-1/2}L^{-3/2}$, where $\bar u$ is the
{\em rms} velocity and $L$ is the integral scale length. This
longstanding universal ${\bar u}^{9/2}$ scaling was confirmed by
the recent Lagrangian experiments \cite{Bodenschatz} to a very
high accuracy, for $R_\lambda >500$. The large scale origin of the
additive noise stochastic intensity, {\em slowly} varying in time,
is precisely the reason of the stretching of exponential tails of
the acceleration probability density function $P(a)$. The observed
Reynolds number dependence of the stretching can be thus
qualitatively understood as the result of increasing coupling of
the acceleration dynamics to large scale dynamics at lower
Reynolds numbers.

In summary, the dynamical equation (\ref{10}) reflects the above
point of view by incorporating exponential of the dynamically
independent $u$, with the prescribed external statistics, as the
intensity of the additive noise. A full dynamical treatment of
Eq.~(\ref{10}) with $u$ viewed as a random function of time can be
made elsewhere. Here, we note only that Langevin type equation
with the additive noise of the form $e^{\omega(t)}L(t)$, with
$\omega(t)$ a longtime correlated stochastic process, was recently
proposed \cite{Mordant} to describe the Lagrangian intermittency
in the context of new multifractal random walk model by Bacry and
Muzy, a continuous extension of discrete cascade models. Our
approach (\ref{10}) meets this result. The difference is that we
approximate the random process $\omega$ to be trivially stationary
and it obeys normal statistics.

In accord to Eqs.~(\ref{5}) and (\ref{10}), the associated
stationary distribution for $F(a)=-a$ is given by
\be\label{Eq:expu}
P(a|u)= C(u)\exp[-e^{u/u_0} a^2/2],
\ee
where $C(u)$ is a normalization constant. From this point of view,
small deviations of the theoretical $P(a)$ given by Eq.~(\ref{1})
from the experimental data \cite{Bodenschatz2} (see
\cite{Aringazin3} for details) could be attributed to small
deviations from Gaussian distribution of the velocity fluctuation
$u$; the flatness was measured to be approximately 2.8 for the
axial and 3.2 for the longitudinal velocity component
\cite{Bodenschatz} compared with the flatness value 3 for an exact
Gaussian distribution. Eq.~(\ref{Eq:expu}) can be used to account
for such deviations in a phenomenological way,
\be\label{Eq:du}
P(a)= \int_{-\infty}^{\infty}du\ C(u)\exp[-e^{u/u_0} a^2/2]g_1(u),
\ee
where $g_1(u)$ is the distribution which is approximately Gaussian
one. However, it should be noted that small departure from the
experimental data can also arise from shortcomings of the
considered one-dimensional Langevin model of developed turbulence.

While it is evident that the three-dimensional Navier-Stokes
equation with a Gaussian random forcing belongs to a class of
non-linear stochastic equations for which one can associate some
generalized Fokker-Planck equations, it is a theoretical challenge
to make a link between the Navier-Stokes equation and such a
surrogate one-dimensional Langevin model, which is, of course, far
from being a full model of the Lagrangian dynamics of fluid in the
turbulent regime. This problem is addressed in the forthcoming
paper \cite{Aringazin5}. Here, we note only that the Lagrangian
description simplifies the problem of finding the stationary
probability measure since the advection term, which dominates in
the developed turbulent flow, is incorporated to the definition of
Lagrangian acceleration. Also, large scale dynamics and an
interaction between the large scale and small scale dynamics
within the inertial range are important to provide understanding
of the origin of noises (multiplicative and additive ones) used in
such Langevin models and a weak dependence of the noise statistics
on the small scale dynamics at high Reynolds numbers.

\section{Conclusions}

(i) The presented approach implies a specific form of the
probability density function $f(\beta)$, given by Eq.~(\ref{7}),
which stems from the assumption that velocity fluctuations are
normally distributed with zero mean, Eq.~(\ref{5}), as it is
confirmed by the experimental data~\cite{Bodenschatz,Bodenschatz2}
for high-Reynolds-number turbulent flows, to a good accuracy.

(ii) Given the exponential dependence (\ref{9}) the presented
formalism leads to a consideration of the log-normal distribution
$f(\beta)$, which was proved to be relevant from the experimental
point of view~\cite{Beck4}. Also, such a distribution is in
agreement with the log-normal distribution of mean energy
dissipation in Kolmogorov 1962 picture. An exponential relation of
the type (\ref{9}) requires a physical treatment in the context of
Langevin, or Fokker-Planck, equation which can be made elsewhere.
Here, we note that the exponential dependence on the velocity
fluctuation, $\exp[u]$, indicates a specific variance of the
absolute value of velocity increment during the time-evolution,
and can be thought of as a kind of Lyapunov instability in the
velocity space.

(iii) One may also be interested in using some other
distributions, instead of the normal one (\ref{6}), to derive
$f(\beta)$ and the associated theoretical distribution $P(a)$.

(iv) Finally, one can try other appropriate functions $B(u)$, for
instance,
\be\label{11}
B(u)=e^{u/u_0}\sum c_ku^k,
\ee
where $c_k$ are constants, instead of (\ref{9}), to reproduce the
experimental data on the acceleration statistics of fluid particle
in turbulent flow. In particular, for $\beta_c=0$ and the function
$B(u)=u^2$ the map doubly covers the interval
$[0,\infty]\ni\beta$, with each covering being a monotonic map,
and we obtain $u=\pm\sqrt{\beta}$ and $|du/d\beta|=\beta^{-1/2}$.
The basic equation (\ref{7}) yields $f(\beta)\sim\beta^{-1/2}
\exp[-\beta/(2s^2)]$, which is $\chi^2$ probability density
function of order one.

Acknowledgement. The authors are grateful to referee for valuable
comments, which allowed to improve physical argumentation in the
revised version of the paper.

\end{document}